\documentclass[aps,pra,floatfix,noshowkeys,epsfig,graphics,natbib]{revtex4}%
\usepackage{graphicx}
\usepackage{amsmath}
\usepackage{amsfonts}
\usepackage{amssymb}

\newcommand{\newc}{\newcommand}
\newc{\beq}    {\begin{equation}}
\newc{\eeq}    {\end{equation}}

\newc{\beqa}    {\begin{eqnarray}}
\newc{\eeqa}    {\end{eqnarray}}
\newc{\bs}    {\section}
\newc{\no}    {\\ \nonumber}

\def\apj{{\em Astrophys. J.  }}
\def\apjl{{\em Astrophys. J. Lett. }}
\def\mnras{{ Mon. Not. Roy. Astron. Soc.  }}

\newc{\st}    {\stackrel}

\begin{document}

\title{ Galaxies with Fuzzy Dark Matter}

\author{Jae-Weon Lee}
\affiliation{ Department of Electrical and Electronic Engineering, Jungwon University,
            85 Munmu-ro, Goesan-eup, Goesan-gun, Chungcheongbuk-do
              28024, Korea}

\begin{abstract}
This is a  brief review on some properties of galaxies in the fuzzy dark matter model, where
 dark matter is an ultra-light scalar particle with mass $m = O(10^{-22})eV$.
 From quantum pressure,  dark matter has a halo length scale  which can  solve the small scale issues of the
 cold dark matter model, such as the core-cusp problem, and explain many other observed mysteries of galaxies.
\end{abstract}

\maketitle

\section{Introduction}

Dark matter (DM) is one of the main ingredients of the universe providing a gravitational attraction to form cosmic structures~\cite{Silk:2016srn}.
The most popular DM model is the  cold dark matter (CDM)  model  in which numerical
simulations successfully reproduce  the observed large-scale structures
 of the universe, such as clusters of galaxies. However, it
encounters some difficulties in explaining  small-scale structures at  galactic scales,
such as the core-cusp problem (predicting the cusped central halo density which is not observed)
and the missing satellite problem (predicting  many more small galaxies than observed)
~\cite{Salucci:2002nc,navarro-1996-462,deblok-2002,crisis}.
Therefore, we need a variant of the CDM  that acts as CDM on a super-galactic scale while
suppressing the formation of smaller structures.
This requires a natural length scale with a small galaxy size on the order of $kpc$.

Recently,  interest  in
the fuzzy DM model as an alternative to the CDM has been revived.
  In this model, DM particles are ultra-light scalars  with mass $m = O(10^{-22})eV$ in a Bose-Einstein condensate (BEC) \cite{2009JKPS...54.2622L,2014ASSP...38..107S,2014MPLA...2930002R,2014PhRvD..89h4040H,2011PhRvD..84d3531C,2014IJMPA..2950074H,Marsh:2015xka,Hui:2016ltb}.
This tiny DM particle mass leads to a very high  DM particle number density,
 which means 
the wave functions
of the particles overlap.  A huge, but finite, length scale related to the Compton wavelength $\lambda_c=1/m\sim 0.1 pc$ of the particles
naturally arises in this model.
Unlike conventional CDM particles that move incoherently, fuzzy DM particles in the BEC state
move collectively and form a coherent wave with the de Broglie wavelength $\lambda_{dB}=O(kpc)>\lambda_c$.
Despite the tiny mass, the fuzzy DM particles are non-relativistic because the particles
 in a BEC move as a heavy single entity.
This model has many other names,  such as BEC DM, scalar field DM, ultra-light axion (ULA), and wave$/\psi$ DM.


The idea that galactic DMs are   condensated ultra-light scalar particles
has been repeatedly suggested  ~\cite {1993ApJ...416L..71W,Schunck:1998nq, PhysRevLett.84.3037,PhysRevD.64.123528,repulsive,fuzzy,
 corePeebles,Nontopological,Mielke2009174,PhysRevD.62.103517,Alcubierre:2001ea,2012PhRvD..86h3535P,2009PhRvL.103k1301S,
 Fuchs:2004xe,Matos:2001ps,0264-9381-18-17-101,PhysRevD.63.125016,Julien,Boehmer:2007um, 2012arXiv1212.5745B,Eby:2015hsq}.
  In Ref. \citealp{1983PhLB..122..221B},
self-gravitating bosons with the de Broglie wavelength of the typical galaxy size
were considered.
In Ref. \citealp{1989PhRvA..39.4207M},
 DM halos  as a ground state of scalar fields  are investigated ~\cite{Matos:2003pe}.
In  Ref. \citealp{Sin:1992bg},  Sin   tried to explain  the observed  flat rotation curves (RCs)
 by using  excited states of the fuzzy DM and obtained the particle mass $m\simeq 3 \times 10^{-23} eV$  by fitting
  the observed RC of  galaxy NGC2998.
 Lee and Koh ~\cite{myhalo} suggested that   DM halos were   giant boson stars
 and considered the effect of  self-interaction.
In this paper, we briefly review some properties of galaxies in the fuzzy DM model.

\section{Fuzzy dark matter and the small scale crisis}
We still lack an exact particle physics model of the fuzzy DM.
 The fuzzy DM field can be a
   scalar field $\phi$
with an action
\beq
\label{action}
 S=\int \sqrt{-g} d^4x[\frac{-R}{16\pi G}
-\frac{g^{\mu\nu}} {2} \phi^*_{;\mu}\phi_{;\nu}
 -U(\phi)],
\eeq
where the  potential is given by
$U(\phi)=\frac{m^2}{2}|\phi|^2$.
 In the Newtonian limit, this action  leads to the following  Schr$\ddot{o}$dinger Poisson equation (SPE), which 
the macroscopic wave function $\psi$ satisfies:
\beqa
\label{spe}
i\hbar \partial_{{t}} {\psi} &=&-\frac{\hbar^2}{2m} \nabla^2 {\psi} +m{V} {\psi}, \no
\nabla^2 {V} &=&{4\pi G} \rho,
\eeqa
where the rescaled field $\psi=\sqrt{m}\phi$,
 the DM mass density $\rho=m|\psi|^2=m^2|\phi|^2$, and $V(\psi)$ is the gravitation potential.
 We used the natural units $\hbar=1=c$.
Note that the SPE can be seen as a non-linear Schr$\ddot{o}$dinger equation, which can have
dispersion-less soliton solutions,
unlike the ordinary  Schr$\ddot{o}$dinger equation.
The ground state of the SPE is one of the solitions.

The SPE has a useful scaling property for numerical studies:
\beq
\{t,r,\psi,\rho,V\}\rightarrow \{\lambda^{-2}t,\lambda^{-1}r,\lambda^{2}\psi,\lambda^{4}\rho,\lambda^{2}V\},
\eeq
where $\lambda$ is a scaling parameter.
This leads to the following scaling law of parameters:
\beq
\{M,E,L\}\rightarrow \{\lambda M,\lambda^{3}E,\lambda L\},
\eeq
where $M$ is the mass, $E$ is the energy, and $L$ is the angular momentum of a dark matter distribution.

For an understanding the role of the fuzzy DM in
the formation of cosmological structures,
reducing the Schr\"{o}dinger equation  to
a fluid equation by using
the Madelung relation ~\cite{2011PhRvD..84d3531C,2014ASSP...38..107S},
$\psi(r,t)=\sqrt{\rho(r,t)}e^{iS(r,t)}$, is useful.
This gives
 an Euler-like equation
\beq
\frac{\partial \textbf{v}}{\partial t} + (\textbf{v}\cdot \nabla)\textbf{v} +\nabla V
+\frac{\nabla p}{\rho} -\frac{\nabla Q}{m} =0,
\label{euler}
\eeq
and a continuity equation
\beq
\frac{\partial \rho}{\partial t} + \nabla \cdot (\rho \textbf{v})=0.
\label{continuity}
\eeq
Here, the fluid velocity $\textbf{v}\equiv \nabla S/2m$
and the quantum potential $Q\equiv{\hbar^2 \Delta \sqrt{\rho}}/({2m\sqrt{\rho}})$.
The pressure  $p$ can come from a self-interaction, if any self-interaction exists.
 (We have ignored  the cosmic expansion for simplicity.).

The quantum pressure
${\nabla Q}/{m}$ from the uncertainty principle is the key difference between the fuzzy DM model and
the conventional   CDM models.
Below the galactic scale, the quantum pressure  suppresses 
 small structure formation, while
 at scales larger than galaxies, the quantum pressure becomes negligible, and the fuzzy
DM behaves like CDM.
This interesting property  makes the fuzzy DM an ideal alternative to the CDM 
because it  resolves
 the small-scale problems of the CDM model while sharing the merits of the CDM~\cite{2009NJPh...11j5029P}.

 To find the length scale during  structure formation
 perturbing the above equations
 around $\rho=\bar\rho$, $\textbf{v}=0$, and $V=0$ is useful.
 One can get the following equation for the density
 perturbation $\delta\rho\equiv \rho-\bar{\rho}$,
 \beq
 \label{pert}
  \frac{\partial^2 \delta\rho}{\partial t^2}+\frac{\hbar^2}{4m^2}\nabla^2 (\nabla^2 \delta \rho)
  -c^2_s \nabla^2 \delta\rho - 4\pi G \bar{\rho}\delta\rho=0,
 \eeq
 where $c_s$ is the sound velocity, and $\bar{\rho}$ is 
 the average density of  background matter \cite{2012A&A...537A.127C, Suarez:2011yf}.

 An equation for the
  density contrast $\delta\equiv\delta \rho/\bar{\rho}=\sum_k \delta_k e^{ik\cdot r}$ with a wave vector $k$,
   \beq
  \frac{d^2 \delta_k}{d t^2} +  \left[(c^2_q+c^2_s)k^2-4\pi G \bar{\rho} \right]\delta_k=0,
 \eeq
  can be derived from the Fourier-transformed version of 
   Eq. (\ref{pert}), where $c_q=\hbar k/2m$ is a quantum velocity.
  Because we can assume $c_s$ to be almost independent of $k$,
   we expect the $c_q$-dependent term from the quantum pressure 
   to dominate only for large $k$ (at a small scale)~\cite{fuzzy,Alcubierre:2002et,PhysRevD.63.063506,Harko:2011jy}.
   This gives
   the quantum Jeans length scale~\cite{1985MNRAS.215..575K,Grasso:1990zg,fuzzy} at a redshift $z$:
\beq
\label{lambdaQ}
\lambda_Q(z)= \frac{2\pi}{k}=\left(\frac{\pi^3 \hbar^2 }{Gm^2\bar\rho(z)}\right)^{1/4}
\simeq 55.6\left(\frac{\rho_b }{m_{22}^2\Omega_m h^2\bar\rho(z)}\right)^{1/4} kpc,
\eeq
where $m_{22}=m/10^{-22}eV$, the Hubble parameter  $h=0.673$, $\rho_b$ is the current matter density, and
 the   matter density parameter
$\Omega_m=0.315$ ~\cite{PDG-2014}.

Interestingly, $\lambda_Q(z)$ determines the minimum length scale of galactic halos formed at $z$
~\cite{Lee:2008ux,Lee:2015cos,Lee:2008jp}. This fact might explain the observed size evolution of the early compact
galaxies ~\cite{Lee:2008ux}. Any perturbation below $\lambda_Q(z)$ decays, and no
 DM structure below this scale can grow.
 This remarkable property  resolves the small scale issues of the CDM model by suppressing the formation
 of too many small structures
 ~\cite{corePeebles,PhysRevD.62.103517,0264-9381-17-13-101,PhysRevD.63.063506}.
The average mass inside $\lambda_Q$ is the quantum Jeans mass
\beq
\label{MJ}
M_J(z)=\frac{4\pi}{3} \bar{\rho}(z) \lambda_Q^3
=\frac{4}{3}
 \pi^{\frac{13}{4}}\left(\frac{\hbar}{G^{\frac{1}{2}}   m}\right)^{\frac{3}{2}} \bar{\rho}(z)^\frac{1}{4},
\eeq
which is the minimum mass
of DM structures forming at z. Therefore, one can expect the minimum mass and size of a galaxy
to have a quantum mechanical origin.

Understanding the core-cusp problem in the fuzzy DM model is easy.
If  no compact object exists at the center of a galaxy, a natural boundary
condition there is a zero-derivative condition, i. e., $\partial \psi/\partial r=0$.
This means the DM density is flat at the center, and this solves the core-cusp problem.
If  a supermassive black hole exists in the DM halo, the central boundary condition should be changed~\cite{UrenaLopez:2002du}.
This property has been argued to explain the M-sigma relation of black holes ~\cite{Lee:2015yws}.
The heavier the black holes are, 
the steeper the slope of the central density profile of DM is. 
This can lead to the M-sigma relation.

\section{Fuzzy dark matter and galaxies}
Because dwarf galaxies are the smallest DM-dominated  objects, they are  ideal
 for studying the nature of DM.
A density perturbation with a size larger than  $\lambda_Q$ can collapse
to form a galactic DM halo.
Therefore, we expect the galactic scale to be smaller than $\lambda_Q$.
 To find a typical size $\xi$ of dwarf galaxies,
we consider the approximate energy function of a spherical DM halo from Eq. (\ref{spe}):
\beq
E(\xi) \simeq\frac{\hbar^2}{2m \xi^2}+\int^{\xi}_0 dr'\frac{Gm}{r'^2}\int^{r'}_0 dr'' 4\pi r''^2
\rho(r''),
\eeq
where $\rho(r)$ is the DM density at $r$.
One can obtain the size of the ground state (the solition),
\beq
\label{xi}
\xi=\frac{\hbar^2}{GMm^2},
\eeq
from
 the condition $\frac{dE(\xi)}{d\xi}=0$ ~\cite{sin1,Silverman:2002qx}.
Here,
$M\equiv \int^{\xi}_0 dr' 4\pi r'^2
\rho(r')$ is the  mass within $\xi$. The size of the halo is inversely proportional to $M$.
A natural assumption  is that
 the quantum Jeans mass is similar to the minimum value of $M$.
 Therefore, the lightest galaxy formed at $z$ has a  typical size
\beq
\label{xiz}
\xi(z)=\frac{\hbar^2}{G M_J(z)m^2}=\frac{3\hbar^{1/2}}{4\pi^{13/4} (G m^2 \bar{\rho}(z))^{1/4}},
\eeq
which has the same form of $\lambda_Q$ (Eq. (\ref{lambdaQ})) with a somewhat smaller constant.
Note that according
to the above equation the lightest (dwarf) galaxy has 
a maximum size among dwarf galaxies in this theory.
 Antlia II was shown to be  close to this upper limit in size
and the velocity of stars is consistent with the theory~\cite{Broadhurst:2019fsl}.
This model seems to  explain the minimum length scale of galaxies  ~\cite{Strigari:2008ib} and
the  size  evolution ~\cite{Lee:2008ux} of the most massive galaxies ~\cite{2009Natur.460..717V}.

On the other hand, an approximate maximum mass of galaxies can  also  be obtained from
the stability condition of boson star theory.
The maximum mass of the ground state is $0.633 m^2_P/m$. If we take this value for
the maximum mass of galaxies ($O(10^{12}M_\odot)$), we can derive a constraint
$m \geq O( 10^{-28}) eV$.
From the maximum stable central density, $m \leq O( 10^{-22}) eV$~\cite{myhalo}.

The finite length scale also implies a finite acceleration scale in this model.
Using an approximation $\partial_r\sim 1/\xi$ in $\nabla Q/m$ in Eq. (\ref{euler}),
 one can obtain a typical
  acceleration scale for DM halos of
\beq
g^\dagger  \equiv \frac{\hbar^2}{2m^2 \xi^3}
=2.2\times 10^{-10} \left(\frac{10^{-22}e{\rm V}}{m}\right)^{2}
\left(\frac{300{\rm pc}}{\xi}\right)^3  \mbox{m/s}^2 ,
\label{gdagger}
\eeq
which is absent in other DM models.
This scale is relevant for
the baryonic Tully-Fisher relation (BTFR)~\cite{Lee:2019ums}, which
is an  empirical relation between the total baryonic mass
of a disk galaxy and its asymptotic rotation velocity.
Interestingly, if we choose the core size of
 the dwarf galaxies ($\sim  300\,\mbox{pc}$ ~\cite{Strigari:2008ib})
 for  $\xi$, we can reproduce the observed
  value $g^\dagger=1.2 \times 10^{-10} \mbox{m/}\mbox{s}^2$ for  Modified Newtonian dynamics (MOND),
  the radial acceleration relation (RAR), and BTFR~\cite{PhysRevLett.117.201101}.
Mond was proposed
to explain the rotation curves without the DM~\cite{1983ApJ...270..365M}.
According to MOND  gravitational acceleration of baryonic matter, $g_b$ should be replaced by
\beq
g_{obs}=\sqrt{g_b g^\dagger},
\label{MOND}
\eeq
when $g_b<g^\dagger$. MOND and RAR may just be effective phenomena of fuzzy DM.


Surprisingly, DM simulations using graphic processing units (GPUs)
with an adaptive mesh refinement (AMR) scheme ~\cite{Schive:2014dra} revealed that
a solitonic core exists in every halo surrounded by granules from DM interference (see Fig. 1).
This configuration is different from the simple excited states of the fuzzy DM or CDM.
An approximate numerical solution can be found in Ref. \citealp{Schive:2014dra}:
\beq
\rho(r)\simeq \frac{\rho_0}{\left(1+0.091(r/r_c)^2\right)^{8}},
\eeq
where the central core density is
$\rho_0=1.9 a^{-1} \left(m/ 10^{-23}eV\right)^2(kpc/r_c)^4 M_\odot/pc^3$
and $r_c$ is the half-density radius.
The outer profile of the halo is similar to the Navarro-Frenk-White (NFW) profile
of the CDM (See Fig. 2.).
The halo mass $M_{halo}$ was also observed to be related to the soliton mass $M$ by
$M\propto M^{1/3}_{halo}$.

\begin{figure}[]
\includegraphics[width=0.5\textwidth]{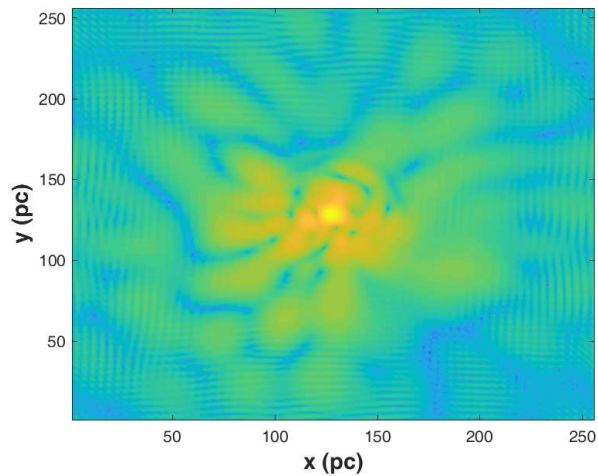}
\caption{(Color online) The two-dimensional section of the three-dimensional
DM density of a model galaxy from our numerical study, which shows a central soliton
and surrounding granules.}
\label{rarfig1}
\end{figure}

\begin{figure}[]
\includegraphics[width=0.5\textwidth]{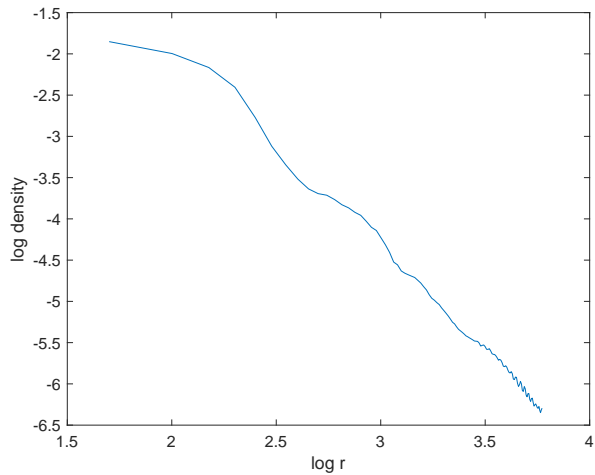}
\caption{ The  DM density versus the radial distance of the model
galaxy shown in Fig.1.}
\label{rarfig1}
\end{figure}


Figure 1 shows  the DM density in a halo in our DM-only numerical simulation using the spectral method.
In this example, 10 small halos with mass $M=10^9 M_\odot$ collide with each other.
From  $\psi$,  one can predict the astrophysical properties of galaxies. For example, the  rotation velocity
at radius $r$ is roughly given by
$
 v_{rot}(r)=\sqrt{\frac{GM(r)}{r}},
$
where $M(r)=4\pi \int^r_0 r'^2 \rho(r') d{r}'$ is the mass within $r$.
This equation can be used to  investigate the RCs of galaxies
 \cite{Matos:2003pe,Lesgourgues2002791,Robles:2012uy,Schive:2014hza} in this model. 
 The mass $m \sim 10^{-22}eV$, which is consistent with other cosmological constraints,
  was obtained by fitting RCs.

\section{Discussions}
Numerical studies in this model so far are mainly DM-only simulations. For a more precise simulation of large galaxies we need to understand
the role of baryon matter such as stars or gas. For example,
it was shown that the gravitational potentials of the fuzzy DM
   induce spiral arm patterns of stars in  galaxies~\cite{2012arXiv1212.5745B}.
In Ref. \citealp{Chan_2018} it was numerically shown that the flat RCs  appear only
when we include visible matter in large galaxies.
 In summary, the fuzzy DM  with mass about $10^{-22}eV$ can explain many mysterious properties of galaxies.
 To find conclusive proofs, we need more precise fuzzy DM simulations with visible matter.

\subsection*{Acknowledgments}
This work was supported by NRF-2020R1F1A1061160.
%

\end{document}